\documentclass[twocolumn,showpacs,amsmath,amssymb]{revtex4}
\usepackage{graphicx}
\begin{document}
\title{Tachyon field inspired dark energy and supernovae constraints}
\author{Jie Ren$^1$}
\email{jrenphysics@hotmail.com}
\author{Xin-He Meng$^{2,3,4}$}
\email{xhm@nankai.edu.cn} \affiliation{$^1$Theoretical Physics
Division, Chern Institute of Mathematics, Nankai University, Tianjin
300071, China} \affiliation{$^2$Department of physics, Nankai
University, Tianjin 300071, China} \affiliation{$^3$BK21 Division of
Advanced Research and Education in physics, Hanyang University,
Seoul 133-791, Korea} \affiliation{$^4$Department of physics,
Hanyang University, Seoul 133-791, Korea}
\date{\today}
\begin{abstract}
The tachyon field in cosmology is studied by applying the generating
function method to obtain exact solutions. The equation of state
parameter of the tachyon field is $w=-1+\epsilon\dot{\phi^2}$, which
can be expressed as a function in terms of the redshift $z$. Based
on these solutions, we propose some tachyon-inspired dark energy
models to explore the properties of the corresponding cosmological
evolution. The explicit relations between Hubble parameter and
redshift enable us to test the models with SNe Ia data sets easily.
In the current work we employ the SNe Ia data with the parameter
$\mathcal{A}$ measured from the SDSS and the shift parameter
$\mathcal{R}$ from WMAP observations to constrain the parameters in
our models.
\end{abstract}
 \maketitle

\section{Introduction}
One of the most challenging tasks in modern cosmology is to
understand the nature of dark energy, which powers the late-time
accelerating expansion of the Universe \cite{rie98,wmap,snls}. The
simplest candidate for dark energy is the cosmological constant, but
it brings serious fine-tuning problem \cite{sw}. However, it is
quite probable that the dark energy with other components in the
Universe is not so simple that it should be described by a more
realistic and complicated EOS \cite{eos1,eos2,eos3,eos4,eos5} or
modified gravity \cite{xh} or scalar field. The scalar field models,
to alleviate the ``old" cosmological constant problem, with a large
variety of potentials have been introduced to study the dynamical
effects of the dark energy (see Ref.~\cite{qu1,qu2} for reviews).
The exact solutions of scalar field models are limited due to the
complicated equations, yet a generating function method is proposed
in Ref.~\cite{sen00,chi99}. Once a generating function is given,
other variables, such as the Hubble parameter and the potential, can
be expressed by some integrations of the generating function. This
useful method can provide us with a large class of exact solutions
of scalar field models.

The tachyon field is originated from the D-brane action in string
theory \cite{sen}. Also, it can be introduced by a simple manner as
follows. The canonical scalar field and the tachyon field can be
regarded as a generalization of the Lagrangian for non-relativistic
and relativistic particles, respectively, \cite{bag02}
\begin{eqnarray}
L_c=\frac{1}{2}\dot{q}^2-V(q) &\rightarrow&
L_c=\frac{1}{2}\partial_\mu\phi\partial^\mu\phi-V(\phi),\\
L_t=-m\sqrt{1-\dot{q}^2} &\rightarrow&
L_t=-V(\phi)\sqrt{1-\partial_\mu\phi\partial^\mu\phi}.
\end{eqnarray}
Although the tachyon field is unstable, it is with an interesting
perspective to explain the dark energy behaviors, which has been
intensely studied in the literature
\cite{ta1,ta2,ta3,ta4,ta5,ta6,ta7}. A natural application of this
framework is to a constant potential $V(\phi)=\sqrt{A}$ for the
Chaplygin gas EOS $p=-A/\rho$, which may give a unified description
of the dark matter and dark energy. However, the exact solutions of
the tachyon field are difficult to obtain. Although the asymptotic
behaviors are studied thoroughly, it is difficult to make some
trustful predictions that can be tested by data. In this paper, we
apply the generating function method into the case of the tachyon
field to solve the the equations and obtain physical quantities.

Recent cosmic observational constraints indicate that the current
EOS parameter $w\equiv p/\rho$ of dark energy is around $-1$,
probably below $-1$, which is called the phantom region
\cite{ph1,ph2} and even more mysterious in the cosmological
evolution stages. The phantom scalar field is introduced by
modifying the kinetic term sign in the Lagrangian to be negative.
Similarly, the phantom tachyon field is also proposed and studied
\cite{pht1,pht2,pht3}. In the present work, we obtain exact
solutions for either the quintessence or the phantom case of the
tachyon field in terms of a given generating function. Inspired by
these results, we propose some effective EOS parameters of dark
energy and employ the SNe Ia data to constrain our models.

The paper is organized as follows: In the next section we present
the general equations and show some examples. In Sec. III we give
exact solutions for some special cases of the generating function.
In Sec. IV we employ the SNe Ia data to constrain dark energy
models. The last section is the conclusion and discussion.

\section{Basic equations and solutions}
We consider the Friedmann-Robertson-Walker metric in the flat space
geometry ($k$=0) as the case favored by recent observational data
\begin{equation}
ds^2=-dt^2+a(t)^2(dr^2+r^2d\Omega_2^2).
\end{equation}
The energy-momentum tensor for the perfect fluid is
\begin{equation}
T_{\mu\nu}=(\rho+p)U_\mu U_\nu+pg_{\mu\nu},
\end{equation}
where $U^\mu=(1,0,0,0)$ is the 4-dimensional velocity in comoving
coordinates. From the Einstein equation
$R_{\mu\nu}-\frac{1}{2}g_{\mu\nu}R=\kappa^2T_{\mu\nu}$, where
$\kappa^2=8\pi G$, we obtain the Friedmann equations
\begin{equation}
H^2=\frac{\kappa^2}{3}\rho,\quad\dot{H}=-\frac{\kappa^2}{2}(\rho+p),\label{eqns}
\end{equation}
where $H=\dot{a}/a$ is the Hubble parameter. The conservation
equation for energy, $T^{0\nu}_{;\nu}=0$, yields
$\dot{\rho}+3H(\rho+p)=0$.

For a canonical scalar field $\phi$, the energy density $\rho$ and
the pressure $p$ are
\begin{equation}
\rho=\frac{1}{2}\dot{\phi}^2+V(\phi),\quad
p=\frac{1}{2}\dot{\phi}^2-V(\phi),
\end{equation}
respectively. For a given potential $V(\phi)$, it is difficult to
find exact solutions due to the nonlinear equations. However, a
class of exact solutions can be obtained in terms of the generating
function $F(\phi)=\dot{\phi}$ \cite{sen00,chi99}. The solutions for
the physical quantities are
\begin{subequations}
\begin{eqnarray}
t(\phi) &=& \int\frac{d\phi}{F(\phi)},\\
H(\phi) &=& -\frac{\kappa^2}{2}\int F(\phi)d\phi,\\
V(\phi) &=& \frac{3}{\kappa^2}H(\phi)^2-\frac{1}{2}F(\phi)^2,\\
a(\phi) &=& a_0\exp\left(\int\frac{H(\phi)}{F(\phi)}d\phi\right),
\end{eqnarray}
\end{subequations}
which are expressed only by some integration forms.

We can give another example. An effective Lagrangian density for the
DBI inflation inspired from the superstring theory is \cite{kin07}
\begin{equation}
\mathcal{L}=-f^{-1}(\phi)\sqrt{1-2f(\phi)X}+f^{-1}(\phi)-V(\phi),
\end{equation}
where $X=-\frac{1}{2}g^{\mu\nu}\partial_\mu\phi\partial_\nu\phi$.
The tension of the brane is $f^{-1}(\phi)=T_3h^4(\phi)$, where
$h(\phi)$ is the warped factor in the metric
\begin{equation}
ds_{10}^2=h^2(r)ds_4^2+h^{-2}(r)(dr^2+r^2ds_5^2).
\end{equation}
In terms of the brane tension $T(\phi)$, the Lagrangian density is
\cite{mar08}
\begin{equation}
\mathcal{L}=-T(\phi)\sqrt{1-\dot{\phi}^2/T(\phi)}+T(\phi)-V(\phi),
\end{equation}
for a homogeneous and isotropic Universe. The corresponding EOS is
\begin{equation}
\rho=(\gamma-1)T(\phi)+V(\phi),\quad
p=\frac{\gamma-1}{\gamma}T(\phi)-V(\phi),
\end{equation}
where the $\gamma$ is the Lorentz factor defined as
\begin{equation}
\gamma\equiv\frac{1}{\sqrt{1-\dot{\phi}^2/T}}.
\end{equation}
In terms of the generating function $F(\phi)=\dot{\phi}$, the
solutions of the Hubble parameter and the potential are given by
\begin{subequations}
\begin{eqnarray}
H &=& -\frac{\kappa^2}{2}\int\big(\gamma(\phi)-\gamma^{-1}(\phi)\big)\frac{T(\phi)}{F(\phi)}d\phi,\\
V &=& \frac{3}{\kappa^2}H^2-\big(\gamma(\phi)-1\big)T(\phi).
\end{eqnarray}
\end{subequations}

\section{Solutions for the tachyon field}
The EOS for the tachyon field $\phi$ is
\begin{equation}
\rho=\frac{V(\phi)}{\sqrt{1-\epsilon\dot{\phi}^2}},\quad
p=-V(\phi)\sqrt{1-\epsilon\dot{\phi}^2},
\end{equation}
where $\epsilon=\pm 1$ and the case $\epsilon=-1$ is for the phantom
tachyon field. We find that the generating function
$F(\phi)=\dot{\phi}$ can also be utilized to the tachyon field case.
The solutions are generally given by
\begin{subequations}
\begin{eqnarray}
t(\phi) &=& \int\frac{d\phi}{F(\phi)},\\
H(\phi) &=& \frac{2}{3\epsilon\int F(\phi)d\phi},\\
V(\phi) &=& \frac{3}{\kappa^2}H(\phi)^2\sqrt{1-\epsilon F(\phi)^2},\\
a(\phi) &=& a_0\exp\left(\int\frac{H(\phi)}{F(\phi)}d\phi\right).
\end{eqnarray}
\end{subequations}
We neglect the integration constant in the above expressions,
therefore, these solutions are only a small class of solutions. The
EOS parameter $w=p/\rho$ thus is given by
\begin{equation}
w=-1+\epsilon\dot{\phi}^2.
\end{equation}
Here we can see clearly that $\epsilon=1$ is corresponding to the
quintessence case, while $\epsilon=-1$ the phantom case. To get
definite answer for physical systems we need exact solutions. In the
following section we pay particular attentions to some special
cases.

\subsection{$F(\phi)=c$}
If $F(\phi)=c$, where $c$ is a constant, the solutions are
\begin{subequations}
\begin{eqnarray}
t(\phi) &=& \frac{\phi}{c},\\
H(\phi) &=& \frac{2}{3\epsilon c\phi},\\
V(\phi) &=& \frac{4\sqrt{1-\epsilon c^2}}{3\kappa^2c^2\phi^2},\\
a(\phi) &=& a_0\phi^{2/(3\epsilon c^2)}.
\end{eqnarray}
\end{subequations}
The integration constants have been ignored for simplicity. The
potential $V(\phi)\propto\phi^{-2}$ has been studied in
Ref.~\cite{ex3,ex4}. It is very natural to obtain this solution by
means of the generating function. The EOS parameter in this case is
$w=-1+\epsilon c^2$. If $c=1$ and $\epsilon=1$, the solution of the
scalar factor therefore is $a(t)=a_0t^{2/3}$, i.e., the tachyon
field behaves like the dust ($w=0$). And if $c$ is close to $0$, it
behaves just like the simplest cosmological constant ($w\sim -1$).
The phantom case contains a future singularity in the cosmological
evolution. If $c=1$, the scale factor behaves as
$a(t)=a_0(t-t_s)^{-2/3}$ when $t\to t_s$. In this case,
$a\to\infty$, $\rho\to\infty$, and $|p|\to\infty$, which means that
the future singularity is the Big Rip \cite{rip}.

\subsection{$F(\phi)=\alpha\phi^\beta$ and $F(\phi)=\alpha e^{\beta\phi}$}
If $F(\phi)=\alpha\phi^\beta$, where $\alpha$ and $\beta$ are free
parameters, the solutions are
\begin{subequations}
\begin{eqnarray}
t(\phi) &=& \frac{\phi^{1-\beta}}{\alpha(1-\beta)},\\
H(\phi) &=& \frac{2(\beta+1)}{3\epsilon\alpha\phi^{\beta+1}},\\
V(\phi) &=& \frac{4(\beta+1)^2\sqrt{1-\epsilon\alpha^2\phi^{2\beta}}}{3\kappa^2\alpha^2\phi^{2\beta+2}},\\
a(\phi) &=&
a_0\exp\left(-\frac{\beta+1}{3\epsilon\alpha^2\beta\phi^{2\beta}}\right).
\end{eqnarray}
\end{subequations}
The solutions for $F(\phi)=\alpha e^{\beta\phi}$ are
\begin{subequations}
\begin{eqnarray}
t(\phi) &=& -\frac{1}{\alpha\beta}e^{-\beta\phi},\\
H(\phi) &=& \frac{2\beta}{3\epsilon\alpha}e^{-\beta\phi},\\
V(\phi) &=& \frac{4\beta^2}{3\kappa^2\alpha^2}e^{-2\beta\phi}\sqrt{1-\epsilon\alpha^2e^{2\beta\phi}},\\
a(\phi) &=&
a_0\exp\left(-\frac{1}{3\epsilon\alpha^2}e^{-2\beta\phi}\right).
\end{eqnarray}
\end{subequations}
If we ignore the integration constants, the EOS parameter contains a
factor $[\ln(1+z)]^{-1}$, which is obviously divergent when $z=0$ as
today. Another example is that the exact solution for the potential
$V(\phi)=4/(3\kappa^2\cos\phi)$ can be obtained by using
$F(\phi)=\sin\phi$. As a more realistic case, we can add a
$\ln(1+z)$ factor in the EOS parameter and test the cosmological
evolution with data.

\section{Supernovae constraints of tachyon-inspired models}
Recent year observations of the SNe Ia have provided the direct
evidence for the cosmic accelerating expansion of our current
Universe. Any model attempting to explain the acceleration mechanism
should be consistent with the SNe Ia data implying results, as a
basic requirement. Recently, lots of relations of the Hubble
parameter $H$ and the redshift $z$ are proposed to test the dark
energy component with observational data, e.g., we have found that
the viscosity without cosmological constant possesses a
$(1+z)^{3/2}$ term contribution to $H$ \cite{eos4}. Technically, the
method of the data fittings is illustrated in Refs.~\cite{fit} for
example.

The EOS parameter $w(z)$ has been obtained in condition that the
tachyon field is the dominated component in the Universe. In the
realistic Universe, dark energy is mixed mainly with the dust
(including ordinary matter and dark matter). The component of dust
contributes a $(1+z)^3$ term to $H^2$, and the dark energy component
contributes a $(1+z)^{3(1+w)}$ term, where $w$ is its constant EOS
parameter, i.e.,
\begin{equation}
H^2=H_0^2[\Omega_\textrm{m}(1+z)^3+(1-\Omega_\textrm{m})(1+z)^{3(1+w)}].\label{eq:H2}
\end{equation}
As an approximation, we assume this addition law for the mixture of
the dust and dark energy is valid if $w$ is with small variations.
See the APPENDIX for details.

We study five models to fit the data for comparison, and the result
is summarized in Table I. The first one is the $\Lambda$CDM model.
The second is $w=-1+w_1$, which is for the case $F(\phi)=c$. For the
third model, we propose an EOS parameter of dark energy as
\begin{equation}
w=-1+w_1\ln(1+z),\label{eq:mod3}
\end{equation}
where $w_1$ is a parameter. The fourth is
\begin{equation}
w=-1+w_1[\ln(1+z)]^{w_2},\label{eq:mod4}
\end{equation}
where $w_1$ and $w_2$ are parameters. We see that the tachyon field
behaves like the variable cosmological constant, thus we expect that
it can be regarded as a possible explanation to the dark energy
behaviors. Moreover, the the fifth model includes an additional
parameter $w_0$ for a more general parametrization
\begin{equation}
w=-1+w_0+w_1[\ln(1+z)]^{w_2}.\label{eq:mod5}
\end{equation}

\begin{figure}[]
\includegraphics{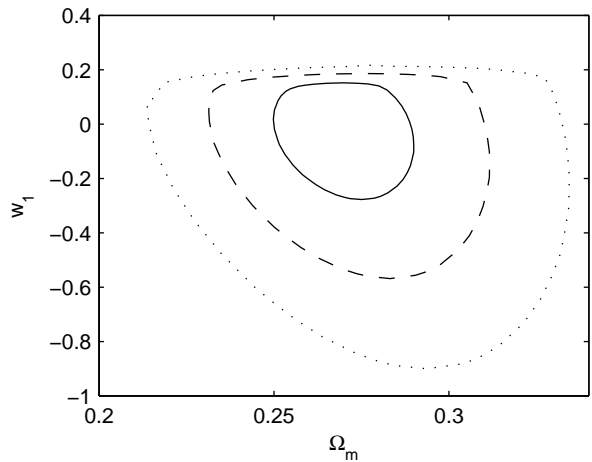}
\caption{\label{fig1} The $1\sigma$ (solid line), $2\sigma$ (dashed
line), and $3\sigma$ (dotted line) contour plots of
$\Omega_\textrm{m}$-$w_1$ relation in model (iii).}
\end{figure}

\begin{figure}[]
\includegraphics{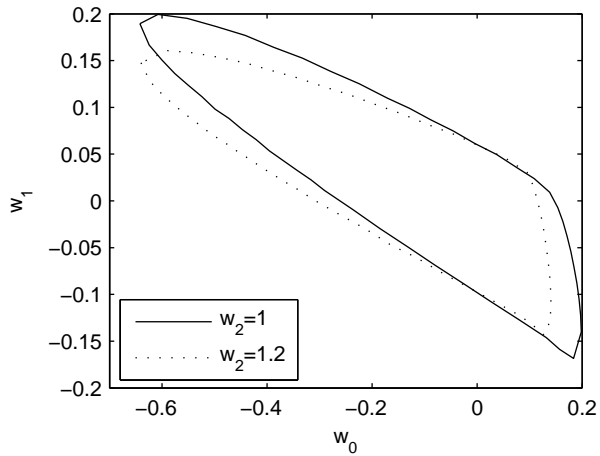}
\caption{\label{fig2} The $1\sigma$ contour plots of $w_0$-$w_1$
relation in model (v) showing no particular preference to
quintessence or phantom.}
\end{figure}

The $\chi^2$ is calculated from
\begin{eqnarray}
\chi^2=\sum_{i=1}^{n}\left[\frac{\mu_{\rm obs}(z_i)-\mathcal{M'}
-5\log_{10}D_{L\rm th}(z_i;c_\alpha)}{\sigma_{\rm obs}(z_i)}\right]^2\nonumber\\
+\left(\frac{\mathcal{A}-0.469}{0.017}\right)^2+\left(\frac{\mathcal{R}-1.716}{0.062}\right)^2,
\end{eqnarray}
where $\mathcal{M'}=\mathcal{M}-M_{\rm obs}$ is a free parameter and
$D_{L\rm th}(z_i;c_\alpha)$ is the theoretical prediction for the
dimensionless luminosity distance of a SNe Ia at a particular
distance, for a given model with parameters $c_\alpha$. The
parameter $\mathcal{A}$ is defined in Ref.~\cite{cala} and the shift
parameter $\mathcal{R}$ in Ref.~\cite{calr}. We will perform a
best-fit analysis with the minimization of the $\chi^2$, with
respect to $\mathcal{M'}$, $\Omega_\textrm{m}$, $w_0$, $w_1$, and
$w_2$ by employing the SNLS data \cite{snls} combined with
$\mathcal{A}$ and $\mathcal{R}$ to constrain our models. For the
model (iii), the $\Omega_m$-$w_1$ relation is plotted in
Fig.~\ref{fig1}, from which we can see that the phantom case of the
tachyon field is slightly favored. For the model (v), the
$w_0$-$w_1$ relation is plotted in Fig.~\ref{fig2} for two
particular choices of $w_2$, with a wide range of possibilities
shown.

\begin{table*}
\caption{\label{tab:t2} Summary of the best fit parameters}
\begin{ruledtabular}
\begin{tabular}{cccc}
& Models by Eq.~(\ref{eq:H2}) & Best fit of parameters & $\chi^2$\\
\hline (i) & $w=-1$
& $\Omega_\textrm{m}=0.269$ & 111.124\\
(ii) & $w=-1+w_1$
& ($\Omega_\textrm{m},w_1)=(0.270,-0.0169)$ & 111.082\\
(iii) & Eq.~(\ref{eq:mod3})
& ($\Omega_\textrm{m},w_1)=(0.270,-0.0234)$ & 111.114\\
(iv) & Eq.~(\ref{eq:mod4})
& ($\Omega_\textrm{m},w_1,w_2)=(0.269,0.00698,2.6309)$ & 111.063\\
(v) & Eq.~(\ref{eq:mod5}) & see Fig.~\ref{fig2} &\\
\end{tabular}
\end{ruledtabular}
\end{table*}


\section{Conclusion and discussion}
We have obtained a class of exact solutions of the tachyon field in
the framework of Friedmann universe. The solutions of the Hubble
parameter $H$, the scale factor $a$, and the potential $V$ are
expressed by general integrations of a given generating function
$F(\phi)$. We list the exact solutions of some special generating
functions, such as the constant, the power-law, and the exponential
functions. By the results one is tempted to postulate that the
tachyon field may provide a possible explanation to the dark energy.
According to the solutions of the tachyon field, we propose some
parameterized EOS parameters of dark energy in the cosmological
evolution, thus these tachyon-inspired dark energy models predict
some new $H$-$z$ relations in cosmology. We employ the SNLS data
with the parameters $\mathcal{A}$ and $\mathcal{R}$ to constrain our
models. The results show that the tachyon field can be a candidate
for the dark energy, and the phantom case is slightly favored to fit
the SNe Ia data, though there is still no way to rule out the
simplest cosmological constant as a good dark energy candidate.

In this work, the tachyon field that causes the late-time
accelerating expansion of the Universe, instead of the problematic
yet economic cosmological constant. The generating function method
enables us to obtain exact solutions to explore the properties of
the tachyon field. Furthermore, another class of scalar field, the
k-essence \cite{kes} has also related to the DBI action and likewise
this generating function method can similarly be applied to it.
Beside, other approaches are also interesting, such as in
Ref.~\cite{od}. Currently, lots of exotic components are proposed to
explain the cosmic dark components (dark matter and dark energy) and
too many of them fit the data well. However, we believe that the
seemingly chaotic situation would be improved as the incoming more
precise data sets in observational cosmology available and we may
gain some new knowledge not far away.

\section*{ACKNOWLEDGEMENTS}
X.H.M. thanks Profs. S. D. Odintsov, I. Brevik, and L. Ryder as well
as Dr. B. Saha for helpful discussions. X.H.M. is supported by NSF
10675062 of China, and BK21 Foundation.

\appendix
\section{A note on the mixture of dust and dark energy}
If there are several components in the Universe, we can solve the
Friedmann equations to obtain the Hubble parameter $H(z)$. Assuming
that only the $i$th component is in the Universe, we can solve a
relation $H_i(z)$. The question is whether the addition law
$H(z)^2=\sum_iH_i(z)^2$ is correct. This is not always true in
general. The reason is as follows. From Eq.~(\ref{eqns}), the
equation determining the Hubble parameter is
$\dot{H}=-\frac{3}{2}H^2+\sum_i\Lambda_i$ if there are the dust and
other components $\Lambda_i$ in the Universe. This equation can be
rewritten as
\begin{equation}
\frac{1}{2}(1+z)\frac{d(H^2)}{dz}=\frac{3\gamma}{2}H^2-\sum_i\Lambda_i.
\end{equation}
If each $\Lambda_i$ is only explicitly dependent on the redshift $z$
or just constant, this is an inhomogeneous linear differential
equation, and $\Lambda_i$ are the inhomogeneous terms. According to
the theory of linear differential equations, the solution of the
this equation is equal to the summation of the solutions when each
nonhomogeneous term exists. Therefore, the conclusion is that if the
component contributes a term which is only explicitly dependent on
the redshift $z$ (equally, the scale factor $a$) or constant, the
addition law is correct.

For example, the curvature term is proportional to $1/a^2$, and the
cosmological constant is a constant. Therefore, the corresponding
terms of the dust, the curvature, and the cosmological constant can
be added to have $H^2$ as a polynomial. In the case of the bulk
viscosity, which contributes a term proportional to $H$
\cite{bre02,eos4}, even an explicit $H(z)$ relation unable to be
obtained, consequently, $H^2$ cannot be separated to a $(1+z)^3$
term and a dark energy term simply. The tachyon field is far more
complicated than the only $z$-dependent term, thus Eq.~(\ref{eq:H2})
may not exactly valid. However, since the $\Lambda$CDM model is in
good agreement with the globally observational data, the behavior
the dark energy should not be deviated from the cosmological
constant too far away. It is reasonable to assume that the addition
law is valid for the mixture of the dust and dark energy as an
approximation, as that is correct for the limit case to the
cosmological constant.

\end{document}